\documentclass[conference]{acmsiggraph}

\usepackage{color}
\usepackage{overpic}
\usepackage{soul}
\usepackage{amsfonts}
\usepackage{wrapfig}
\usepackage{multirow}
\usepackage{amssymb, amsmath, amsfonts}
\usepackage{color}
\usepackage[ruled,vlined,onelanguage,linesnumbered]{algorithm2e}

\usepackage[T1]{fontenc}

\newcommand{\mypara}[1]{\vspace{0.05in} \noindent \textbf{#1.}}

\SetCommentSty{algocomment}

\newcommand{\vnudge}{\vspace{-.1in}}
\newcommand\mbf[1]{\mathbf{#1}}

\graphicspath{{./figures/}}

\definecolor{turquoise}{cmyk}{0.65,0,0.1,0.1}
\definecolor{purple}{rgb}{0.65,0,0.65}
\definecolor{dark_green}{rgb}{0, 0.5, 0}
\definecolor{orange}{rgb}{0.8, 0.2, 0.2}
\definecolor{red}{rgb}{1, 0, 0}

\title{Towards Zero-Waste Furniture Design}

\author{Bongjin Koo\thanks{Joint first authors} \\ University College London \and 
Jean Hergel$^*$ \\ INRIA Nancy \and
Sylvain Lefebvre  \\  INRIA Nancy \and
Niloy J. Mitra  \\ University College London}
\pdfauthor{}


\begin{document}

 \teaser{
   \includegraphics[width=\textwidth]{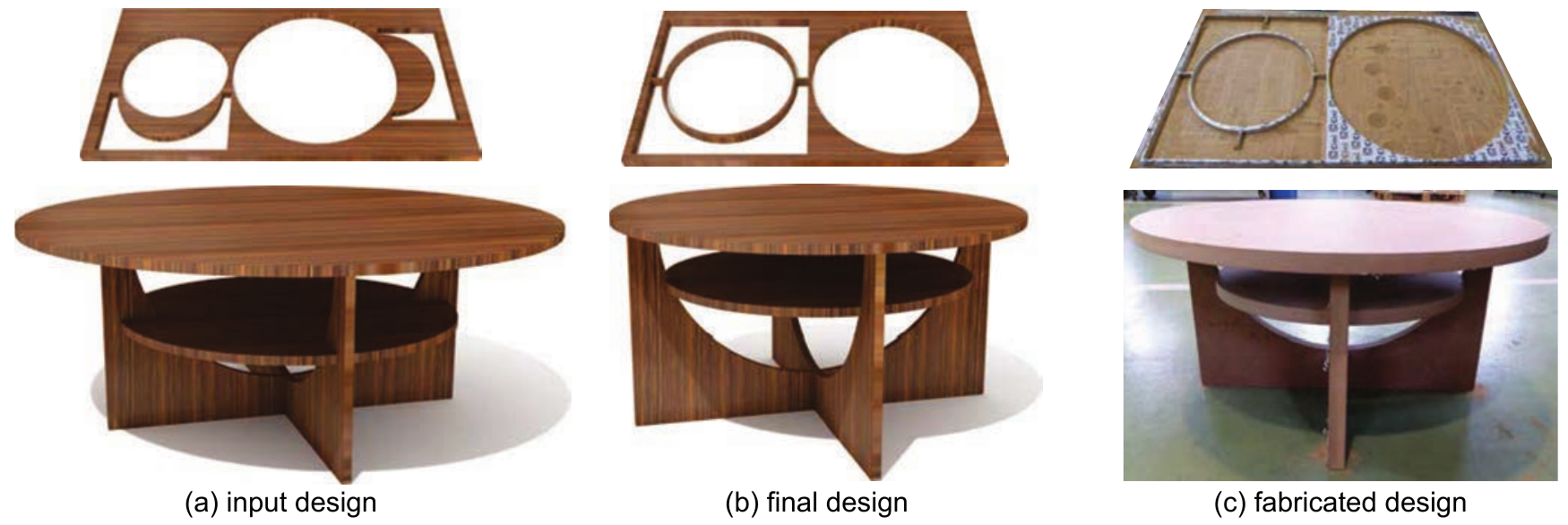}
   \caption{We introduce waste-minimizing furniture design to dynamically analyze an input design~(a) based on its 2D material usage~(see inset) and design specifications to assist the user through (b)~multiple design suggestions to reduce material wastage~(see inset). The final user design can directly be exported for laser cutting and be assembled~(c). In this case, wastage was reduced from $22\%$ to $11\%$. 
   }
   \label{fig:teaser}
   \vnudge
 }

\maketitle

\begin{abstract}

In traditional design, shapes are first conceived, and then fabricated.
While this decoupling simplifies the design process, it can result in inefficient material usage, especially where 
off-cut pieces are hard to reuse. 
The designer, in absence of explicit feedback on material usage remains helpless to effectively adapt the design -- even though design variabilities exist.
%
%
%In this paper, 
We investigate {\em waste minimizing furniture design} wherein based on the current design, the user is 
presented with design variations that result in more effective usage of materials. 
Technically,  we dynamically analyze material space layout to determine {\em which} parts to change and {\em how}, while maintaining original design 
intent specified in the form of design constraints.
We evaluate the approach on simple and complex  furniture design scenarios, and demonstrate effective material usage that is difficult, if not impossible, to achieve without computational support.

\if0
%
%Materials have a strong influence on what shapes can and cannot be easily fabricated.
In traditional design, shapes are first conceived, and then fabricated.
%
%This is a very simplistic assumption in case of materials that are cut and put together, rather than molded or deposited into shape.
%
This decoupling of  design and fabrication 
leads to inefficient material usage, especially in cases of hard-to-reuse materials in the context of 
laser cut fabrication.
%
The designer, in absence of usable knowledge about future material usage, remains helpless to effectively adapt the design.
%
%Wastage is further amplified when multiple boards are used, or the same design is replicated multiple times. 
%
In this paper, we investigate how the current design along with its material usage can be  analyzed to suggest users
 how to effectively refine the designs. 
 %Effectively, the analysis guides the users to better conform to material usage.
%
Technically,  we enable this by dynamically analyzing material space layout and design effectiveness (e.g., sagging, space usage) in order to generate suggestions to support guided form finding at interactive rates.
%
We evaluate the approach on simple and complex  furniture design scenarios, and demonstrate effective material usage that is difficult, if not impossible, to achieve without computational support.
%
\fi

\end{abstract}

\if0
\begin{CRcatlist}
  \CRcat{I.3.5}{Computer Graphics}{Computational Geometry and Object Modeling}{Geometric algorithms, languages, and systems}
\end{CRcatlist}
\fi

\keywordlist

\TOGlinkslist

%\copyrightspace

% ----------------------------------------------------------------
\section{Introduction}

Furniture design is an exercise in form-finding wherein the designer arrives at a final form by balancing aesthetics, object function, and cost. 
Typically, design variations are manually explored by a mixture of guesswork, prior experience, and domain knowledge. 
Without appropriate computational support, such an exploration is often tedious, time consuming, and can result in wasteful choices.

In furniture manufacturing, both for mass production and for customized designs, material wastage plays a deterrent role. This not only leads to increased production cost (typically 5-15\% wastage due to off-cuts), 
but also hampers ongoing efforts towards green manufacturing~\cite{dain:09}. For an extensive report, please refer to the guideline from the British Furniture Manufacturer~\cite{bfm:03}. 
Hence, there has been a growing interest in {\em zero-waste furniture} in an effort to reduce material wastage. A notable example being Maynard's `Zero-waste Table.' 
Computational support for designing such waste-reducing furniture, however, is largely lacking. 

Material considerations are typically appraised only {\em after} a shape has been designed. 
While this simplifies designing, it leads to unnecessary wastage: at design time, the user can at best guess to account for how the shape will be physically realized, and can easily fail to effectively adjust the design to improve material utilization.
% p

In recent years, algorithms have been developed to economically 3D print given designs. 
For example, approaches have been proposed to cleverly breakup a given shape into parts that better pack together in print volumes~\cite{Luo:2012:CPM,Vanek2014,Chen2015,Yao2015}, adaptively hollow shape interiors to save print materials~\cite{Stava:2012:SRI:2185520.2185544,Prevost:2013:MSB:2461912.2461957,Wang:2013:CPO:2508363.2508382,Dumas:2014:BGA:2601097.2601153}, explore parameter space variations for manufacturable forms~\cite{Shugrina2015}, or 
design connector geometry to remove the need for any secondary connector parts~\cite{Fu2015}. 
However, improving material utilization by explicitly allowing design changes has been less studied.

In this work, we introduce the problem of {\em waste-minimizing furniture design}, and investigate it in the context of flatpack furniture design~(cf., \cite{flatpack:06}) using laser cut wooden parts. Specifically, we study the interplay between furniture design exploration and cost-effective material usage. %Note that by cost-effective, we refer to reducing material wastage and making efficient use of fabrication machine cycles.)
By directly coupling the two, we empower the users to make more informed design decisions. 
%Hence, both 2D material layout (i.e., how the planks are arranged and cut from the master board) and 3D design influence the final shape. 
%
%Thus, the user actively changes the design with the system guidance to adapt the design to reduce material wastage. 
Note that this is fundamentally different from locking a designed shape, and then trying to best fabricate it.

For example, in Figure~\ref{fig:teaser}, the user starts with an initial concept indicating  {\em design constraints} (e.g., symmetry, desired height, etc.). Our system analyzes {\em material usage} by computing a dynamic 2D layout of the parts and proposes design modifications to improve material usage without violating specified design constraints (i.e., design intent). 
Note that such adaptations are often in the form of synchronous movement of multiple parts affected by both design and material layout considerations, which are difficult to mentally imagine.
The user can select any of the suggestions, either in its entirety or in part. She can further update the set of design constraints by locking parts of the current design, and the process continues. Thus, the user scopes out a design space via constraints, and our algorithm refines the design to reduce material wastage while restricting changes to the indicated design space. 
%
%More generally, our system supports simultaneously designing $M$ objects spanning across $N$ boards (see also supplementary video).
%
%Functional constraints (e.g., placed weights), when user-provided, are also analyzed to suggest appropriate design modifications (see also supplementary video).

Technically, we achieve the above by using the current material layout to {\em dynamically} discover a set of relevant layout constraints.
The algorithm has a discrete aspect involving {\em which} part to change based on the current 2D layout, and a continuous aspect involving {\em how} to adapt the part attributes based on the current material space layout without violating user-specified design constraints.
Even for a fixed design, exploring the space of all possible packing is a combinatorial NP-hard problem. Instead, we locally analyze 
a set of candidate packings to determine which parts to modify and how to change them to optimize material utilization. We demonstrate that by dynamically analyzing a set of current packings, we can efficiently and effectively couple the 2D layouts and the constrained 3D designs. The user is then presented with different waste-reducing design variations.

We evaluated the system to create a variety of simple and complex designs, and fabricated a selection of them. We also performed a user study with both designers and novices to evaluate the effectiveness of the system. The performance benefits were particularly obvious in case of complex designs involving different design constraints. 
In summary, we:
\begin{itemize}
\item introduce the problem of material waste minimizing furniture design; and
\item propose an algorithm that dynamically analyzes 2D material usage to suggest design modifications to improve material usage without violating user-specified constraints. 
%\item evaluate the system on a variety of examples and fabricate a selection of the designs.
\end{itemize}

% ----------------------------------------------------------------
\section{Related Work}

%Decades of research have been devoted to facilitate creation of novel shape designs, rationalize them for targeted fabrication methodologies, and  evaluate design effectiveness. However, how the three aspects interact to influence the final design has not been explicitly investigated.

\mypara{Material considerations} Physical materials play an important role in 3D printing an object. Various approaches have been developed to economically and efficiently produce a designed object. For example, adaptively hollowing out interiors and adding struts to create durable yet cost-effective 3D printouts~\cite{Stava:2012:SRI:2185520.2185544}, cleverly hollowing the shape interiors in conjunction with shape deformation to ensure stability of the final shape~\cite{Prevost:2013:MSB:2461912.2461957}, or perform FEM analysis to decide wall thickness and parameters to ensure model endurance under known or unknown forces~\cite{Zhou:2013:WSA:2461912.2461967,Lu:2014:BSW:2601097.2601168}. Techniques for designing scaffolds, both interior~\cite{Wang:2013:CPO:2508363.2508382} and exterior~\cite{Dumas:2014:BGA:2601097.2601153}, have been developed for cost-effective 3D printing by reducing wastage. 
Hu et al.~\shortcite{Hu2015} propose to optimize the shape of a 3D model to reduce support structures used during 3D printing. Alternatively, methods have been developed to decompose and pack 3D models for reducing assembly cost, support material, printing time or making big objects printable on small 3D printers~\cite{Luo:2012:CPM,Vanek2014,Yao2015}. Dapper~\cite{Chen2015} also employs a decompose-and-pack approach for minimum assembly cost, support material and build time when using 3D printers. It breaks 3D objects into pyramidal primitives, then finds good packing configurations to achieve the goal. 

In the context of laser cut fabrication, Hildebrand et al.~\shortcite{Hildebrand2012} and Schwartzburg and Pauly~\shortcite{Schwartzburg13} explore how to rationalize a given design for fabrication out of planar sheets. Further, material wastage has been investigated by testing  various packing strategies from computational geometry community (cf., \cite{jylanki2010thousand}) to efficiently layout the parts in the material space. More recently, Saakes et al.~\shortcite{Saakes:2013:PMC:2501988.2501990} proposed an interactive system to allow the user to interactively layout parts for more personalized usage. Such methods, however, do not explicitly {\em modify} the original designs in order to improve material usage. 

\mypara{Fabrication-aware design}
Recently, the growing popularity of personalized fabrication has motivated researchers to develop algorithms to adapt existing shapes to make them better suited for physical construction. Examples include abstracting shapes as a collection of slices to be laser cut~\cite{msm_slices_siga_11,Hildebrand2012,Schwartzburg13,CPMS14},
as foldable popups~\cite{li2010popup,li2011a},
developing toolkit to allow user to draft directly using a handheld laser pointer to control high-powered laser cutters~\cite{Mueller:2012:ICI:2380116.2380191},
computationally designing gear trains to support part movement for converting animated characters to working physical automata~\cite{Coros:2013:CDM}, introducing necessary joint geometry to create non-assembly articulated objects~\cite{Bacher2012,Cali12ThreeDee}, or supporting an example-driven fabrication paradigm~\cite{Schulz2014}. 
To simplify fabrication, Fu et al.~\shortcite{Fu2015} suggest a method to generate a globally-interlocking furniture assembly that enables easy disassembly/reassembly of furniture, {\em without} using glue, screws, etc. 
%
%An interesting alternative is a data-driven approach~\cite{Schulz2014} to author a design. The interactive system supports combining parts from a collection of example templates to form new models. 
%
Such methods, however, are chiefly used to adapt existing shapes {\em after} they have been designed, rather than to guide the user to refine the designs to reduce material-wastage.

\mypara{Guided design} In the context of exploratory design, Xu et al.~\shortcite{xu_sig12} proposed a fit-and-diverse framework to allow users to interactively guide model synthesis and exploration, while Talton et al.~\shortcite{Talton:2009:EMC} exposed a parameterized shape space for model creation. These efforts, however, focus on aspects of digital content creation without fabrication and material considerations. Recently, Shugrina et al.~\shortcite{Shugrina2015} developed a system that allows novices to easily customize parametric models while maintaining 3D-printability of the models. In a work closely related to our motivation, Umetani et al.~\shortcite{uim_guidedExploration_sigg12} use stability and durability of materials to propose design modifications, thus computationally guiding the users. With a similar motivation, we investigate the impact of material usage  in the context of guided design. We are unaware of prior attempts investigating how material usage can be analyzed to refine the designs.

\if0
\mypara{Topology optimization for modeling}
Topology optimization~\cite{sigmund.13.smo} considers optimizing
shapes under specific objectives, such as maximizing rigidity or reaching a
target deformation, and under constraints regarding the material usage.
In the field of computer graphics these techniques have been recently adapted
by Christiansen et al. for design purposes~\shortcite{topoOptSimplicital:14},
and for balancing 3D models~\shortcite{autobalancing:14}.
We propose a method to optimize for the rigidity of our designs which is
inspired by topology optimization, but further considers the design and layout
constraints required to model furniture.
\fi

\mypara{Constraint-based modeling} In the CAD community, constrained-based modeling (cf., \cite{constrainedModeling:98}) has long been demonstrated as a powerful parametric way to design shapes and interact with them. In the case of existing models, an inverse
analyze-and-edit paradigm has been recently proposed to first discover the constraints present in shapes, and then allow interactive editing~\cite{gsmc_iwires_sig_09,Xu:2009:JMD:1531326.1531341,ZFDOT10}. Such approaches differ on how model parts are abstracted (e.g.,
feature curves, model parts,
or abstracted segments as primitives) and how the inter-part constraints are conformed to. However, these methods have primarily focused on designing shapes for the virtual world where material and fabrication constraints are irrelevant, and hence ignored.

% ----------------------------------------------------------------
\section{Design Workflow}
\label{sec:overview}

Our goal is to propose design variations that minimize material wastage without violating original design intent. In this section, we present the proposed system as experienced by the user, and describe the main algorithmic details in the subsequent sections. Here we particularly focus on how the user encodes her design intent.

The user starts by choosing the desired material (i.e., thickness of wooden planks) and the number and dimensions of the master board(s).
Our system considers rectangular master boards --- in practice these can represent new boards or left over rectangular spaces in already used boards.
The user starts by loading an initial part-based 3D object design, either created in a modeling system or as a parameteric model. The parts can be rectangular or have curves boundaries. The user also indicates a set of  
{\em design constraints}. In our implementation, we support: equal length (e.g., $l_i=l_j$), sum of lengths (e.g., $l_i+l_j+\dots = l_k+\dots$), fixed length (e.g., $l_i=c$),
equal position, symmetric parts, ground touching, and coplanarity among indicated planks.
%(Coplanarity, if present, are also auto-detected templates among touching planks}.
%
%
The user can additionally specify that the object should fit an indicated volume (e.g., in between two walls) and the internal space in the form of inner volume indicating minimal shelf dimensions. 

\begin{figure}[h!]
\centering
\begin{overpic}[width=.8\columnwidth]{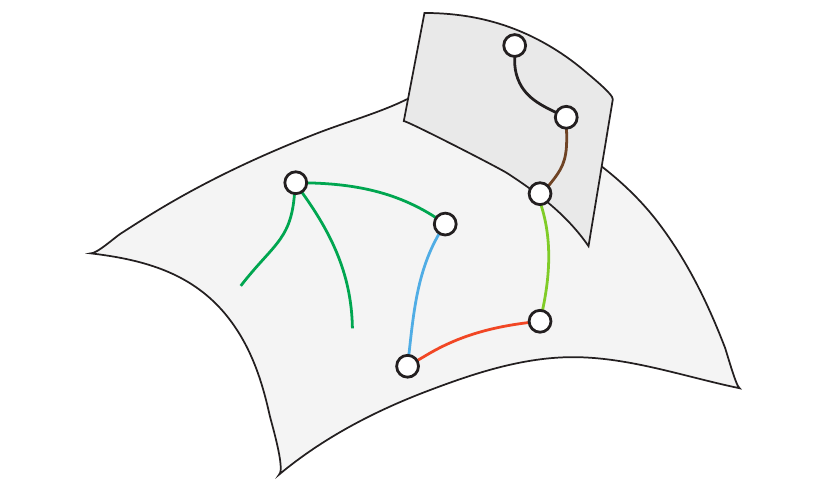}
\put(34,5){$\mathcal{S}_1$}
\put(52,51){$\mathcal{S}_2$}

\put(32,32){\small $M_1$}
\put(46,29){\small $M_2$}
\put(43.5,11){\small $M_3$}
\put(67,17){\small $M_4$}
\put(60,32){\small $M_5$}
\put(70,44){\small $M_6$}
\put(63.5,50){\small $M_7$}
\end{overpic}
\caption{Our algorithm discovers  design variations in shape space. The user starts from a design $M_1$ along with indicated design constraints, and the algorithm 
seeks for wastage minimizing variations by interleaving between topologically different material layouts (indicated by changes in curved paths) or continuous changes to the layouts (indicated by same colored curves). For example, paths $(M_i,M_j)$ denote  
continuous design changes, while points $M_i$ denotes designs where new layouts are explored (i.e., branch points). The user can switch to another shape space by picking an updated set of design constraints (shape $M_5$ here). Note that 
by construction $M_5$ belongs to both shape spaces $\mathcal{S}_1$ and  $\mathcal{S}_2$.  See Algorithm~\ref{alg:global}. }
\label{fig:shape_space}
\end{figure}

The algorithm suggests multiple design variations that all satisfy the design specifications but achieve different material usages. We measure material usage based on the fraction of the master board(s) utilized. 
The top  suggestions are presented as thumbnails. If the user mouse-overs any thumbnail, the system animates the proposed design modifications. The user can preview the  object- and material-space views, and select her preferred design suggestion. Note that each thumbnail effectively
represents a design exploration path pursued by the algorithm. We provide a slider to move along this path, which is particularly useful for making incremental updates to the design (see Figure~\ref{fig:shape_space} and Section~\ref{sec:algo_overview}).

The user either selects a suggested design variation, or picks part configurations from a suggested shape as additional design constraints (e.g., user can lock the proposed sizes of certain planks). Thus, effectively the user appends or updates the current set of specified design. Note that the new constraints are 
trivially satisfied by the current design, which is critical for subsequent design space exploration (e.g., $M_5$ is in both shape spaces $\mathcal{S}_1$ and  $\mathcal{S}_2$).

%[Phase 2: production]
Once satisfied with a design, she requests for the cutting patterns. She can investigate the design, the material space usage and the cutting patterns, and send the patterns directly for laser cutting.

% ----------------------------------------------------------------

\begin{figure*}[t!]
\centering
\includegraphics[width=\linewidth]{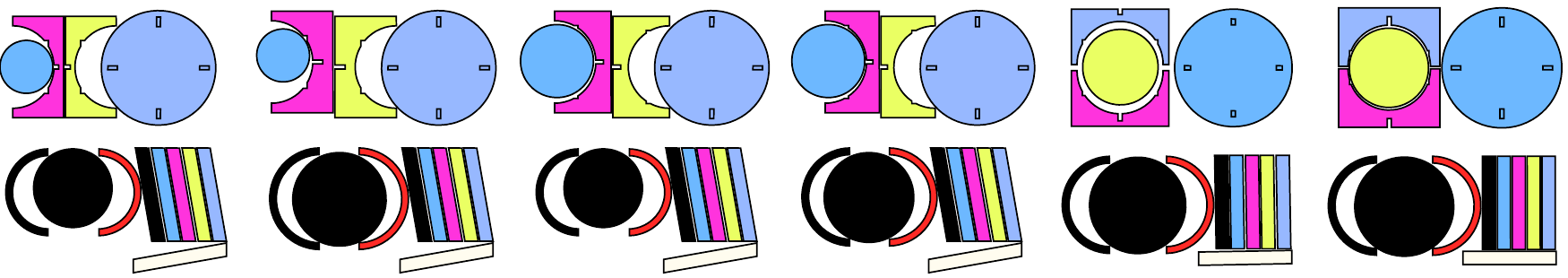}
\caption{Evolution of shape variation across a run of our algorithm on the coffee-table~(top) and low-chair~(bottom) models.}
\label{fig:alog_steps}
\vnudge
\end{figure*}

\section{Overview}
\label{sec:algo_overview}

Our goal is to analyze aspects arising from material considerations, and investigate how design changes affect such considerations. Specifically, we ask how to adapt a furniture design so that it makes better utilization of material in the resultant design layout.
Note that this is the inverse of the design rationalization problem, i.e., instead of taking a design as fixed and best fabricating it, we adapt the design so that the resultant rationalization makes better utilization of available material.
First, we introduce some notations.

\subsection{Parameterized designs}
The design is considered as a function $D(\mbf{X})$ that produces the geometry of a fixed number of parts,
given a configuration vector $\mbf{X}$. The parts can be assembled into a final furniture design. 

We make no assumption as to how $D$ is implemented -- we demonstrate in Section~\ref{sec:results} applications 
using both constrained based furniture design
and parametric designs modeled by CSG. We however expect a continuous behavior
from $D(\mbf{X})$, i.e., small changes in $\mbf{X}$ result in small changes 
in the part shapes. Parametric modelers generally offer such continuity to 
smoothly navigate the space shape. 
% We also input min and max bounds $\mbf{X}_{min}, \mbf{X}_{max}$ for the set of parameters $\mbf{X}$. 

During wastage optimization our algorithm will change the value of $\mbf{X}$ so as to explore
whether changes in part shapes reduce wastage. 
Since we focus on laser cut furniture construction, we assume the parts to have the same thickness $\tau$.
The parts are thus represented as planar polygonal contours extruded orthogonally.

The geometry of a part $p_i$ lies within a bounding box which we represent 
by a six dimensional vector encoding the box center $\mbf{p}_i$ and the lengths of its 
three sides $l^x_i,l^y_i,\tau$ -- the Z axis being aligned with part thickness by convention. 

\subsection{Material space}

Since we focus on laser cut furniture, any 3D design given by a configuration vector $\mbf{X}$ is realized as a layout (i.e., cutting plan) in the material space.
Material space is characterized by the largest {\em master board} that the machine can possibly cut, a rectangle of size $W \times H$. 
In this space, each part $i$ is associated with a position $(u_i,v_i)$ and an orientation
$o_i \in \{0, {\pi}/{2}, {\pi}, - {\pi}/{2}\}$. 

We use $w_i, h_i$ as extent of a part bounding box in the material space along
the x- and y-axis, respectively.
The part box lengths in material space are given by the two plank dimensions other
than thickness. For a plank $i$, of orientation $o_i$, we get one of the two cases:
\[
\begin{array}{lcll}
o_i = 0, o_i=\pi &\Rightarrow& w_i = l^x_i & h_i = l^y_i\\
o_i = -\pi/2, o_i=\pi/2 &\Rightarrow& w_i = l^y_i & h_i = l^x_i\\
\end{array}
\]

The material space positions and orientations are variables in the layout optimization algorithm,
alongside the design parameters $\mbf{X}$ (see Section~\ref{sec:pack}).

When wastage is not a concern and a design easily fits within material space, 
the variables $(u_i,v_i,o_i)$ are independent of the design, i.e., they simply 
adapt to changes in part sizes.
However, as we seek to maximize utilization of the material space, the material
space variables become tightly coupled with the design parameters.
Our layout optimizer therefore jointly optimizes for material space variables
and design parameters to minimize wastage (see Section~\ref{sec:pack})

We next discuss what makes a desirable layout from the point of view
of furniture fabrication.

% -----------------------------------------------------
\subsection{Properties of a good design layout}
\label{sec:pack:cost}

% \syl{Note: this is rewritten with a slightly different approach. The machine fixes max dimensions, other than that we can source
%	any rectangular master board. Is this optimistic? I surely can from the local hardware store, but they cut them at no fee from very large boards. }
Rectangular master boards can be sourced in a large choice of sizes and thicknesses from resellers. 
Therefore, our goal is to achieve a full utilization of rectangular spaces, so that the user can use boards of exactly
the right size and minimize wastage. The machine dimensions determine the maximum extent of a single board.

We measure wastage as the fraction of the space not utilized by the design in its material space bounding rectangle.
Ideally, we want to achieve full utilization, i.e., null wastage.
%
% Note that there can be one or more master boards as required by the design. 
% In the most general setting, our system supports $M$ designs covering $N$ master boards.

\begin{figure}[h!]
	\vnudge
	\includegraphics[width=\columnwidth]{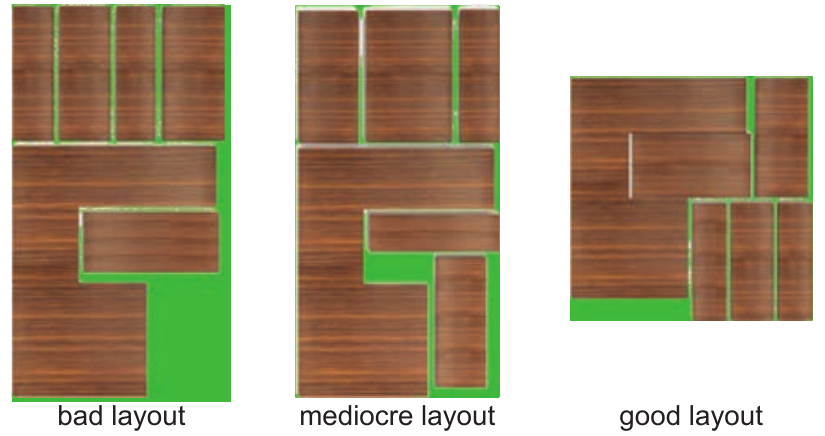}
	\caption{Examples of stages of layout refinement, from bad to mediocre to good. A good layout is characterized by
		less area of material wasted (shown in green). }
	\label{fig:design_layout_goodBad}
\end{figure}

An ideal packing is one that tightly packs all the parts to perfectly fill up one or more rectangular master boards (like a puzzle).
Our system helps the user achieve this by automatically exploring changes improving material space usage (see Figure~\ref{fig:design_layout_goodBad}).
% \syl{and edit the shape of the parts to further minimize wastage.}

\if0
Minimizing the total cost would make little sense as this
would tend to make the design vanish. Instead, we seek to minimize the cost of wastage (see Figure~\ref{fig:design_layout_goodBad}), i.e., the wasted area and cut time.
\fi

\if0
We define the {\em wasted area} as the ratio between the area of the design
and the area of its bounding rectangle. Indeed, a design with a simple rectangular
outer boundary will be very simple to fit in a plank of given size, making its
fabrication and reuse of extra material much simpler. Note that such a requirement is quite characteristic for materials like wood, which cannot be melted and recast like glass or plastic.

{\em Cut time} is minimized by sharing plank boundaries that can then be cut
with a single laser line.
\fi

%-----------------------------------------------------------------
%
% Main algo first
% - change text to describe main algorithm first, re-order
% - text format in algo textsc
% - write co-planar, intuition at least

\section{Design Layout Optimization}
\label{sec:pack}

\begin{algorithm}[b!]
\KwIn{Design function $D$, starting design parameters $\mbf{X}_s$}
\KwOut{Set of best layouts found $\mathcal{L}$}
$O_s \leftarrow $ identity ordering \tcp*[l]{1,2,3,...}
$\mathcal{X} \leftarrow \{(\mbf{X}_s,O_s)\}$\;
\For{$G$ iterations}{
	\ForEach{$(\mbf{X},O) \in \mathcal{X}$}{
		$\mathcal{O} \leftarrow$ \textsc{ExploreOrderings($\mbf{X},O$)}\;
		\ForEach{$O \in \mathcal{O}$}{
		  $\mathcal{X} \leftarrow \mathcal{X} \cup$ \{(\textsc{ImproveDesign}($\mbf{X},O$),$O$)\}\;
		}
    }
	$\mathcal{X} \leftarrow $ \textsc{KeepBests}(K,$\mathcal{X}$)\;
}
$\mathcal{L} \leftarrow \emptyset$\;
\ForEach{$(\mbf{X},O) \in \mathcal{X}$}{
	$\mathcal{L} \leftarrow \mathcal{L} \cup $ \textsc{Docking}($D(\mbf{X})$,$O$)\; 
}
\Return ($\mathcal{L}$)\;
\caption{\textsc{MinWastage}}
\label{alg:global}
\end{algorithm}

The wastage of a layout depends essentially on two factors. The first factor is the quality 
of the packing that can be achieved, given a fixed set of design parts. The second factor is
the set of parts itself, which can be changed through the design parameters $\mbf{X}$.

In our approach we pack the parts using a deterministic docking algorithm that always produces
the same result for a same ordering of the design parts. 
Therefore, a first optimization variable is the order in which the parts are sent to the docking algorithm. 
The second optimization variable is the vector of design parameters~$\mbf{X}$. 
These two variables have different natures: finding
an ordering is a combinatorial problem while the design parameters can be continuously explored.

We therefore proceed in two main steps, first determining a set of good orderings
that then serve as starting points for continuously evolving the design, reducing wastage.
The overall approach is described in Algorithm~\ref{alg:global}. The subroutine \textsc{ImproveDesign}
is described in Section~\ref{sec:improve} while \textsc{ExploreOrderings} is described in Section~\ref{sec:orderings}.
The process restarts for a number of iterations (we use $G=3$) to jump out of local minima reached by the 
continuous design exploration. This results in the shape space exploration illustrated in Figure~\ref{fig:shape_space}.
The process returns the $K$ best found layouts and designs and presents them to the user in thumbnails.
She can then select her favorite design, and if desired update the constraints and restart the exploration
from this point --- which simply calls \textsc{MinWastage} again.
%

% ----------------------------------

\begin{algorithm}[b!]
	\KwIn{Starting design parameters $\mbf{X}$ and ordering $O$}
	\KwOut{Modified design parameters $\mbf{X}_b$ with reduced wastage}
	$L \leftarrow $ \textsc{Docking}(D($\mbf{X}$),$O$)\;
	$\mbf{X}_b \leftarrow \mbf{X}, L_b \leftarrow L$ \;
	$\mbf{X}_c \leftarrow \mbf{X}, L_c \leftarrow L$ \;
	\For{$N$ iterations}{
		$\mbf{X}_b,L_b \leftarrow$\textsc{GrowParts}($\mbf{X}_b,L_b$,$\mbf{X}_c,L_c,O$)\; \label{alg:improve:grow}
		$\mbf{X}_c \leftarrow$\textsc{ShrinkParts}($\mbf{X}_b,L_b$)\; \label{alg:improve:shrink}
		$L_c \leftarrow $ \textsc{Slide}($L_b$,$D(\mbf{X}_c)$)\;
		\tcp{Check for improvement over current.}
		\If{$W(L_c) <  W(L_b)$}{
			$\mbf{X}_b = \mbf{X}_c$, $L_b=L_c$\;
		}
	}
	\Return ($\mbf{X}_b$)\;
	\caption{\textsc{ImproveDesign}}
	\label{alg:improve}
\end{algorithm}

\mypara{Bitmaps}
During optimization we regularly call the parameterized design function $D(\mbf{X})$ to obtain 
a new set of parts after changing parameters. The layout optimization represents parts internally as
bitmaps: each part contour is rasterized at a resolution $\tau$, typically $0.5$ mm per pixel.
This enables fast manipulation of the parts within the layout.
Each part thus becomes a bitmap having either $1$ (inside) or $0$ (outside) in each pixel. The
size of the bitmap matches the part extents in material space $w_i$ and $h_i$. Every time the
design is refreshed a new set of bitmaps is computed for the parts.
The master board is similarly discretized into a regular grid of resolution $\tau$.

% We compute multiple starting points in parallel by launching several threads 
% from different random permutations of the part ordering.

% In this section, we describe how an initial design is laid out in the material space leading to a set of dynamically generated layout constraints, and how the constraints are analyzed to create design refinements. 
% Essentially, we tightly couple the 3D object space and 2D material layout space to computationally generate design refinements leading to reduced material wastage.

% -----------------------------------------------------

\subsection{Design optimization for wastage minimization}
\label{sec:improve}

The design optimization improves the design parameters $\mbf{X}$ to minimize wastage in the layout,
keeping the docking ordering fixed. 
It appears as the subroutine \textsc{ImproveDesign} in Algorithm~\ref{alg:global}.
The pseudo-code for this step is given in Algorithm~\ref{alg:improve}.
Our objective is to suggest design changes that reduce wastage, progressively improving the 
initial layout. The algorithm performs a guided local search by changing the parts -- through the
design parameters -- to reduce wastage.

Prior to considering which parts to modify, we have to answer two questions: First,
how to drive the design parameters $\mbf{X}$ to change only a given part (Section~\ref{sec:sizing}).
This is achieved by relying on the gradients of the part size with respect to $\mbf{X}$. 
Second, we have to decide on how to evolve the layout when parts are changed (Section~\ref{sec:slide}).
We rely on a sliding algorithm that avoids jumps in the layout configuration, thus
producing only small changes in the wastage function when small changes are applied to 
the part sizes.

\mypara{Overall strategy}
Our approach changes the size of parts iteratively with two different steps
in each iteration: \textit{grow} (line~\ref{alg:improve:grow}) and \textit{shrink} (line~\ref{alg:improve:shrink}). 
These steps progressively 
modify the design and keep track of the design of smallest wastage encountered so far. 

The grow step (Section~\ref{sec:grow}) attempts to enlarge the parts so as to reduce wastage. Each 
part is considered and its size is increased for as long as the growth further 
reduces wastage.
When no further improvement can be obtained, we create further opportunities
by shrinking a set of parts (Section~\ref{sec:shrink}). However, randomly shrinking 
parts would be inefficient, as most parts would grow back immediately to their
original sizes. Other parts are tightly coupled to many others in the design $D$,
and shrinking these would impact the entire design. 
Therefore, we analyze the layout to determine which parts have a higher probability
to result in wastage reduction.

% At the end of the process we either return the design, or use it as a new starting
% point to recursively call the design exploration process. The overall approach is
% described in Algorithm~\ref{}, while the next sections detail each component.

\subsubsection{Changing part sizes}
\label{sec:sizing}

During design space exploration the algorithm attempts to vary the part sizes $w_i$ and $h_i$ individually. 
% The part sizes are defined as the maximal extent of the part along the material space dimensions, $w_i$ and $h_i$. 
These dimensions vary as a function of design parameters $\mbf{X}$. In the remainder we use $\mbf{s}(\mbf{X})$ to
designate the vector of all part sizes assembled such that $s_{2i} = w_i$ and $s_{2i+1}=h_i$.

Let us denote $\lambda$ the change of size desired on $s_i$.
Our objective is to compute a design change $\Delta$ such that $s_i(\mbf{X}+\Delta) = s_i(\mbf{X}) + \lambda$.
We denote the vector of changes as $\Lambda = \mbf{s}(\mbf{X}+\Delta) - \mbf{s}(\mbf{X})$. In this process only 
the size $s_i$ should change with others remain unchanged whenever possible, 
that is $\Lambda_{s_j, j \neq i} = 0$ and $\Lambda_{s_i} = \lambda$.

Parts are not independent in the design and therefore there is no trivial link between $\mbf{X}$ and $s_i(\mbf{X})$.
We therefore analyze the relationship through the gradients
$\frac{\partial s_i(\mbf{X})}{\partial x_j}$. These are computed by local finite differencing (depending
on the design analytical expressions may be available). 
Each non-null gradient indicates that parameter $x_j$ influences $s_i$. Multiple parameters may 
influence $s_i$ and parameters typically also influence other variables: there 
exists $k \neq i$ such that $\frac{\partial s_k(\mbf{X})}{\partial x_j} \neq 0$.

To compute $\Delta$ we formulate the following problem. 
Let us consider the components of $\Delta = (\delta_0, ..., \delta_{|\mbf{X}|-1})$.
The change in part sizes due to $\Delta$ can be approximated in the first order 
through the gradients as 
$\mbf{\Lambda} = \sum_i{\delta_i \cdot \frac{\partial \mbf{s}(\mbf{X})}{\partial x_i}}$.
We solve for $\Delta$ such that $\Lambda_{s_i} = \lambda$ and $\Lambda_{s_j,j\neq i} = 0$.

If there are less parameters than part sizes, the problem is over-constrained and 
solved in the least-square sense, minimizing $||\mbf{\Lambda} - (0,...,\lambda,...,0)||^2$. 
If there are more parameters than part sizes, the problem is under-constrained and
solved in the least-norm sense, minimizing $||\Delta||$. We rely on a QR decomposition
of the system matrix to solve for both cases, accounting for possible rank deficiencies
due to overlapping parameters in $\mbf{X}$.

We implement this process as a subroutine \textsc{ChangePartSize($\mbf{X}$,$s_i$,$\lambda$)}, 
with $\mbf{X}$ the current design parameters, $s_i$ the part size to change and $\lambda$ the change to apply.
It returns the new design parameters $\mbf{X}+\Delta$. A second subroutine
\textsc{ChangePartSizes($\mbf{X}$,$\Lambda$)} allows to change the size of multiple parts at once.

\begin{algorithm}[t!]
	\KwIn{current layout $C=(u_0,v_0, ...)$ and set of changed parts $parts$}
	\KwOut{updated layout $L$}
	$L \leftarrow \emptyset$\\
	\ForEach{part $p_i \in parts$ in docking order} { \label{alg:slide:iter}
		\For{N iterations} { \label{alg:slide:num}
			$\Delta_x \leftarrow - smallestLeftFreeInterval(L,p_i)$\; \label{alg:slide:x0}
			\If {$\Delta_x = \emptyset$} { 
				$\Delta_x \leftarrow smallestRightDecollision(L,p_i)$\;
			}  
			$pos_x \leftarrow (u_i + \Delta_x,v_i)$ \; \label{alg:slide:x1}
			$\Delta_y \leftarrow - smallestBottomFreeInterval(L,p_i)$ \; \label{alg:slide:y0}
			\If {$\Delta_y = \emptyset$} {
				$\Delta_y \leftarrow smallestTopDecollision(L,p_i)$ \;
			}	       
			$pos_y \leftarrow (u_i,v_i + \Delta_y)$ \; \label{alg:slide:y1}
			\If{$pos_x = \emptyset$ and $pos_y = \emptyset$} {
				\tcp*[l]{cannot fit masterboard}
				return $\emptyset$ \tcp*[l]{$W(\emptyset) = 1$} \label{alg:slide:failed}
			} 
			\If{$pos_x = pos$ and $pos_y = pos$} {
				break\;
			} 
			\If{$A(box(L \lhd_{pos_x} p_i) < A(box(L \lhd_{pos_y} p_i))$}{ \label{alg:slide:check0}
				$(u_i,v_i) \leftarrow pos_x$
			} \ElseIf{$A(box(L \lhd_{pos_x} p_i) > A(box(L \lhd_{pos_y} p_i)$}{
			$(u_i,v_i) \leftarrow pos_y$	       
		} \Else {
		\If{$\Delta_x<\Delta_y$ and $|\Delta_x| > 0$}{
			$(u_i,v_i) \leftarrow pos_x$
		} \Else {
		$(u_i,v_i) \leftarrow pos_y$ 
	}\label{alg:slide:check1}
}	       
}
$L \leftarrow L \lhd_{(u_i,v_i)} p_i$
}
\Return ($L$)\;
\caption{\textsc{slide}}
\label{alg:slide}
\end{algorithm}

\subsubsection{Updating layouts by sliding}
\label{sec:slide}

As the shapes and sizes of the parts change the layout has to be updated. One option
would be to restart the docking process after each change. However, for a small change
the docking process can produce large discontinuities in the wastage function. 
This makes a local search difficult. Instead, we propose to rely on a sliding 
operation that attempts to continuously update the position of the parts after each change. 
Note that performing such an update while optimizing for a given objective (i.e. wastage)
is a very challenging combinatorial problem, as each part can move in four directions 
(left/right/top/bottom) and multiple cascading overlaps have to be resolved. 
We propose a heuristic approach that works well for small changes in the part shapes. 

\begin{figure}[htb]
	\vnudge
	\centering
	{\includegraphics[width=\columnwidth]{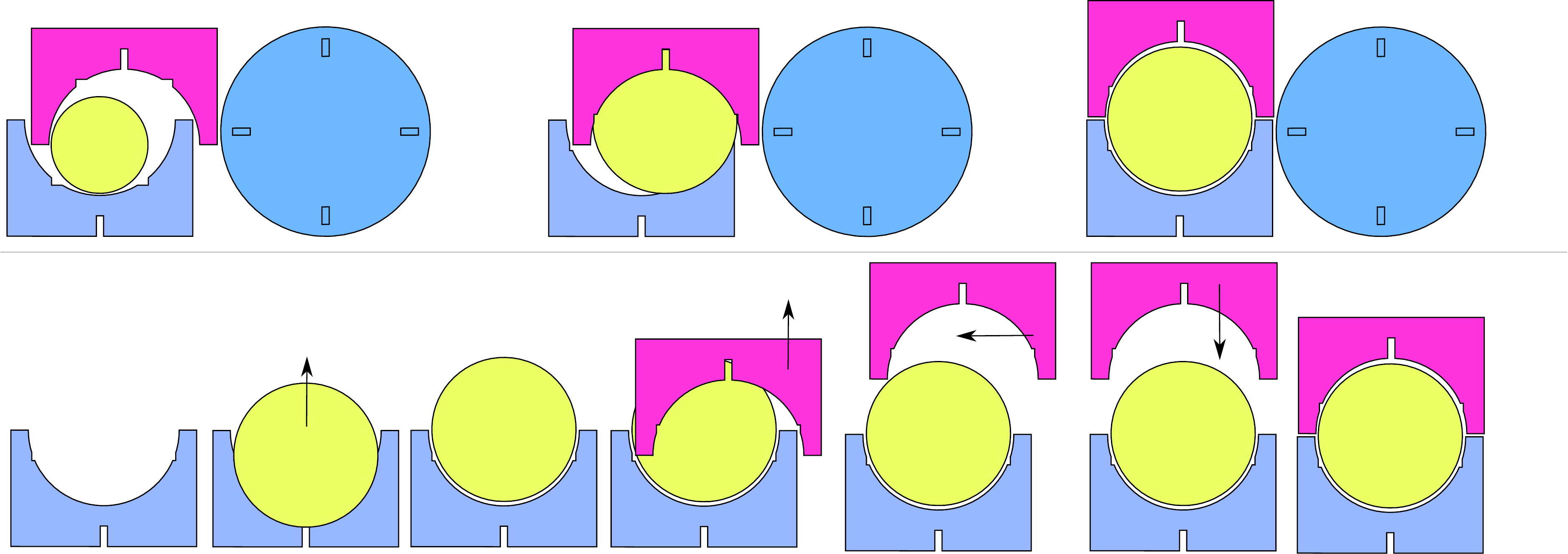}}
	\caption{Sliding a layout after a change of part sizes. 
		\textbf{Top:} From left to right, initial layout, same after change revealing overlaps, layout after sliding.
		\textbf{Bottom:} Moves performed on the three first parts during sliding.
	}
	\label{fig:slide}
\end{figure}

The algorithm is based on the following principle. After changing the part shapes, we 
reintroduce them in an empty layout in order of docking. However, each time a part is reintroduced it may
now have empty space to its left/bottom or it may overlap with previously placed parts.
Both cases can be resolved by a single horizontal or vertical move. However 
a single move is generally not desirable as empty space may remain along the other direction. 
We therefore perform a limited sequence of horizontal/vertical moves. At each
iteration we select between vertical or horizontal by favoring moves that result
in the smallest layout bounding box. In case of a tie, we favor moves to the
left/bottom versus displacements to the top/right.
This is illustrated in Figure~\ref{fig:slide}.

The pseudo-code is given in Algorithm~\ref{alg:slide}. 
In the algorithm we denote by $L$ the layout  
and denote by $L \lhd_{pos} p_i$ the layout obtained when adding 
part $p_i$ at position $pos$ in the master board grid of $L$. $A(.)$ measures the area, $box(L)$ is the bounding rectangle of the layout.
The algorithm iterates over all parts
in docking order (line~\ref{alg:slide:iter}). It then performs a fixed number of sliding
operations on each part (line~\ref{alg:slide:num}) -- we use $N=4$ in our implementation.
Lines~\ref{alg:slide:x0}-\ref{alg:slide:x1} compute a horizontal move, favoring moves
to the left that collapse newly created empty spaces. 
Lines~\ref{alg:slide:x0}-\ref{alg:slide:x1} similarly compute a vertical move. 
Lines~\ref{alg:slide:check0}-\ref{alg:slide:check1} decide whether to select a horizontal move
$pos_x$ or vertical move $pos_y$.

The process may fail if parts can no longer fit in the masterboard. This can happen
either because there is not enough remaining area, or because sliding cascades in large moves
that prevent further insertion of parts. In such cases we return an empty layout which by
convention has a wastage of $1$ (worst possible), line~\ref{alg:slide:failed}.

\subsubsection{Grow step}
\label{sec:grow}

%(line~\ref{alg:grow: })

The grow step is described in Algorithm~\ref{alg:grow}.
The algorithm iterates over all parts in random order (line~\ref{alg:grow:foreach})
and progressively increases the size of a part in a loop (line~\ref{alg:grow:loop}).
Note that the first iteration of the loop determines the starting wastage for
growing this part (lines~\ref{alg:grow:maxw} and~\ref{alg:grow:cmp0}-\ref{alg:grow:cmp1}). 
The process continues until the growth results in an increased wastage (line~\ref{alg:grow:break}).

After each change of parameters the design parts are recomputed (line~\ref{alg:grow:slide}, $D(\mbf{X_e})$) and sliding is
called to adapt the current layout to the change.
The result is checked. If wastage decreases the process continues (line~\ref{alg:grow:cmp1}). 
If not, we first attempt to dock the parts again (line~\ref{alg:grow:dock}). This 
can help continue the growth in cases were sliding fails to resolve overlaps by continuous changes.
If wastage still not improves we stop the growth of this part size (line~\ref{alg:grow:break}).
% Note that the sliding algorithm may fail, returning the max wastage
% which is then ignored line~\ref{alg:grow:cmp0}.

\begin{algorithm}[t!]
	\KwIn{Best design parameters $\mbf{X}_b$ and layout $L_b$ so far, 
		  current design parameters $\mbf{X}_c$ and current layout $L_c$ being explored, ordering $O$.}
	\KwOut{New best design and packing.}
	$improvement \leftarrow true$\;
	\While{$improvement$}{
		$improvement \leftarrow false$\;
		\ForEach{part size $s_i$ in random order} {	\label{alg:grow:foreach}
			$W_e \leftarrow 1$ \tcp*[l]{max wastage} \label{alg:grow:maxw}
			$\mbf{X}_e \leftarrow \mbf{X}_c$,
			$L_e \leftarrow L_c$ \;
			\tcp{Grow a first time and then continue as long as it improves.}
			\While{ true }{ \label{alg:grow:loop}
				$\mbf{X}_e \leftarrow$ \textsc{ChangePartSize($\mbf{X_e}$,$s_i$,$1$)} \tcp*[l]{+1 pix.}
				$L_e \leftarrow $ \textsc{Slide}($L_e$,$D(\mbf{X_e})$)\; \label{alg:grow:slide}
				\If{$W(L_e) > W_e$}{  
					$L_e \leftarrow $ \textsc{Docking}($D(X_e)$,O)\; \label{alg:grow:dock}
				}
				\If{$W(L_e) < W_e$}{  \label{alg:grow:cmp0} % \tcp*[l]{true on first iter} 
				  	$W_e = W(L_e)$\; \label{alg:grow:cmp1}
				} \Else {
					break\; \label{alg:grow:break}
				}
			}
			\tcp{Check for improvement over current.}
			\If{$W_e <  W(L_c)$}{
			  	$\mbf{X}_c = \mbf{X}_e$, $L_c=L_e$\;
			  	$improvement \leftarrow true$\;
			}
	    } 
    }
	\tcp{Check for improvement over global best.}
	\If{$W(L_c) < W(L_b)$}{
		$\mbf{X}_b = \mbf{X}_c$, $L_b=L_c$\;
	}
	\Return ($\mbf{X}_b,L_b$)\;
	\caption{\textsc{GrowParts}}
	\label{alg:grow}
\end{algorithm}

\subsubsection{Shrink step}
\label{sec:shrink}

The goal of the shrink step is to create further opportunities for design changes when
no parts can further grow. 
The typical situation is that a subset of parts are forming \textit{locking chains} between 
respectively the left/right and top/bottom borders.
The parts belonging to these chains prevent any further growth. 
We therefore detect locking chains and select the parts to shrink among these. This often
results in a change of aspect ratio of the masterboard, and new opportunities for other parts
to grow.

The overall approach is described in Algorithm~\ref{alg:shrink}.
It first determines which parts to shrink by calling \textsc{SelectPartsToShrink} and then
computes a change of parameters using the approach described in Section~\ref{sec:sizing}.

\begin{algorithm}[b!]
\KwIn{Best design parameters $\mbf{X}_b$ and layout $L_b$ so far.}
\KwOut{Shrunk design parameters.}
$\mbf{X}_s \leftarrow \mbf{X}$\;
$\mathcal{S} \leftarrow$ \textsc{SelectPartSizesToShrink}($L_b$)\;
$\Lambda \leftarrow (0,...,0)$\;
\ForEach{$s_i \in \mathcal{S}$ }{
	$\Lambda_i \leftarrow -1$ \tcp*[l]{-1 pixel}
}
$\mbf{X}_s \leftarrow$ \textsc{ChangePartSizes($\mbf{X_s}$,$\Lambda$)} \tcp*[l]{-1 pixel}
\Return ($\mbf{X}_s$)\;
\caption{\textsc{ShrinkParts}}
\label{alg:shrink}
\end{algorithm}

The core component is the \textsc{SelectPartSizesToShrink} subroutine, described in Algorithm~\ref{alg:shrink:select}.
The selection starts by gathering all contacts between parts in the layout -- this is done efficiently in the 
discretized layout grid. We first draw the part images into the grid and then check pairs of neighbors
belonging to different parts. This produces the set of left/right and bottom/left contacts between
part sizes (the involved part size is deduced from the part orientation and the considered axis). 
The contacts are oriented from right to left (respectively top to bottom). We similarly
detect which parts touch the borders.
The contact detection is implemented in the \textsc{GatherContactsAlongAxis} subroutine.

Once the contacts are obtained we start from the right (respectively top) border and form locking chains.
Starting from the border, we produce the set of chains iteratively. Each chain $c$ is a sequence
$(left,s_{first}, ..., s_{last})$. At each iteration the chain spawns new chains for each contact pair
$(s_{last},s_{next})$ obtained by augmenting $c$ as $(left,s_{first}, ..., s_{last}, s_{next})$. 
Potential cycles are easily detected as repetition of a same part in the chain and are ignored.
The locking chain computation is implemented in the \textsc{FormContactChains} subroutine.

We next randomly select part sizes to shrink until all locking chains are removed.
The selection probability of each part is designed to avoid too large a jump in the design space.
To achieve this we consider two factors.
First, we compute the number of occurrences of each part in the locking chains, $occ(p_i)$. A part with
many occurrences is a good candidate as shrinking it will resolve multiple locking chains at once.
Second, we seek to avoid shrinking part sizes that are tightly coupled with others in the design $D$. 
We compute the dependence of a part size by counting the number of non-zero entries in the
$\mbf{\Lambda}$ vector computed internally by \textsc{ChangePartSize($\mbf{X_e}$,$s_i$,$-1$)}.

We select part sizes with the following random process. First, we select a number of occurrences $o$
with probability $P(o) = \frac{\sum_{p_i, occ(pi)=o}{occ(o)}}{\sum_{p_i}{occ(p_i)}}$. Then, 
among the parts such that $occ(p_i) = o$ we select a part size $s_i$ with probability 
$P(s_i|occ(s_i)=o) = 1 - \frac{dep(s_i)}{\sum_{p_i, occ(p_i)=o}{dep(p_i)}}$. This process is 
implemented by the \textsc{DrawPartSizeWithProbability} subroutine.

After each part size selection we update the set of locking chain by removing all chains where 
the part size appears.

\begin{algorithm}[t!]
	\KwIn{A layout $L$.}
	\KwOut{Set of part sizes to shrink.}
	$\mathcal{K} \leftarrow \emptyset$\;
	\ForEach{axis $a \in \{X,Y\}$}{
	  $\mathcal{C} \leftarrow $\textsc{GatherContactsAlongAxis}($a$) \;
	  $\mathcal{K} \leftarrow \mathcal{K}~\cup~$\textsc{FormContactChains}($\mathcal{C}$) \;
    }
    $\mathcal{S} \leftarrow \emptyset$\;
    \While{$\mathcal{K} \neq \emptyset$ }{
    	$s_i \leftarrow $ \textsc{DrawPartSizeWithProbability}($\mathcal{K}$)\;
    	$\mathcal{S} \leftarrow \mathcal{S}~\cup~\{s_i\}$\;
    	$\mathcal{K} \leftarrow \mathcal{K}~\setminus $ \textsc{KilledChains}($\mathcal{K}$,$s_i$)\;
    }
	\Return $\mathcal{S}$\;
\caption{\textsc{SelectPartSizesToShrink}}
\label{alg:shrink:select}
\end{algorithm}

% -----------------------------------------------------

\subsection{Exploring orderings}
\label{sec:orderings}

The subroutine \textsc{ExploreOrderings} in Algorithm~\ref{alg:global} performs
a stochastic search of orderings resulting in low wastage layouts.
The process starts from a random order and iteratively considers possible improvements by 
swapping two parts. 
At each iteration, we perform a swap and recompute a layout using the docking algorithm.
If wastage is reduced the swap is accepted, otherwise it is rejected. We apply the process 
for a number of iterations and keep the best ordering 
found as the starting point. We use $|D(\mbf{X})|^2$ iterations, where $|D(\mbf{X})|$ is the 
number of parts.
For each ordering, we use a fast docking algorithm to compute a layout with low wastage.

\mypara{Docking algorithm}
The docking algorithm places each part in order by 'dropping' the next part on 
the current layout either from the right, or from the top. It locally searches for 
the best placement of each part, according to a criterion that minimizes wastage. 
The result is a layout $L$ including all parts.

% We denote by $W(L)$ the wastage of the layout, defined by the ratio of 
% occupied area divided by the area of the bounding rectangle of $L$ \syl{redundant?}.

Given the layout so far our algorithm searches for the best orientation and 
best position for the next part.
We denote by $L_{i-1}$ the layout obtained for the $i-1$ first parts, and
by $L_i \leftarrow L_{i-1} \lhd_{pos} p_i$ the layout obtained by adding the next
part at position $pos$. The docking position $pos$ is computed from 
a drop location $(s,x,o)$, with $s \in \{top,right\}$, $x$ a position along 
the corresponding axis and $o \in \{0,\pi/2,\pi,-\pi/2\}$ an orientation.

\begin{figure}[t!]
	\vnudge
	\centering
	{\includegraphics[width=0.41\columnwidth]{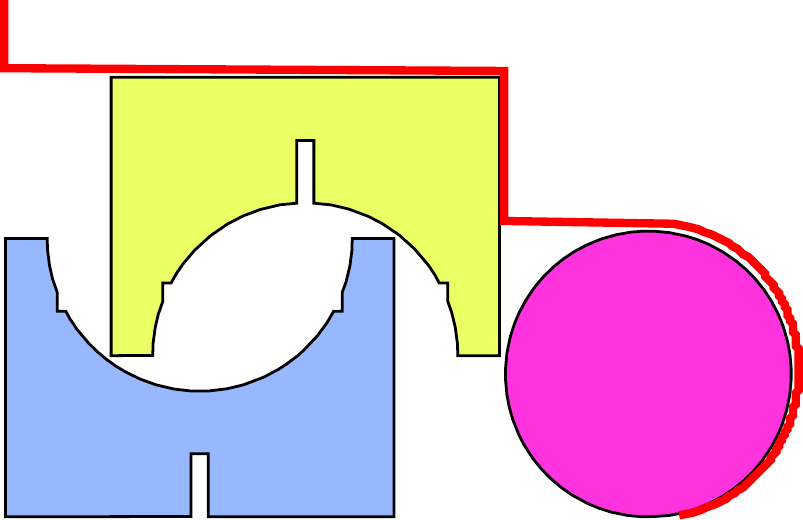}}
	\hfill
	{\includegraphics[width=0.45\columnwidth]{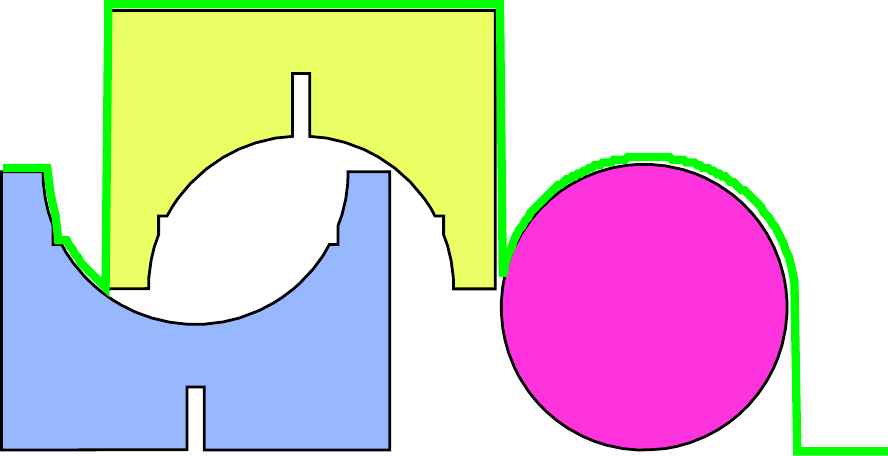}}
	\caption{Height-fields of the layout used to position the next part. \textbf{Left:} Height-field for dropping
		parts from the right (red curve). \textbf{Right:} Height-field for dropping parts from above (green curve).
		These height-fields are maintained every time a new part is added to the layout, and used for fast computation
		of the docking positions. Similar height-fields are pre-computed for the left/bottom of the parts.
	}
	\label{fig:hfields}
\end{figure}

The pseudo code for the docking algorithm is given in Algorithm~\ref{alg:dock}.
The drop locations are ranked according to a {docking criterion} 
that we denote $D(L_{i-1},p_i,pos)$, explained next.
The docking positions are computed from the drop locations by 
the \texttt{ComputeDockingPosition} subroutine.
It is efficiently implemented by maintaining the right/top height-fields of the
current layout as illustrated in Figure~\ref{fig:hfields}. % The left/bottom height-fields of the parts are pre-computed. 
Whenever evaluating a drop location we use the height-fields 
to quickly compute the docking positions that bring the part in close contact with 
the current layout.
% The complexity of this step only depends on the vertical (respectively horizontal) size of the parts height-fields, not on their areas.

\begin{algorithm}[h!]
	\KwIn{Set of parts $P$, order $O$, master board dimensions $W \times H$}
	\KwOut{A layout $L$}
	\ForEach{part $p_i \in P$ following order in $O$} {
		$best \leftarrow \emptyset$ \;
		$bestscore \leftarrow 1$ \;
		\ForEach{drop location $(s,x,o)$} {
			$pos$ $\leftarrow$ \texttt{ComputeDockingPosition}($p_i,(s,x,o)$) \;
			$score \leftarrow D(L_{i-1},p_i,pos)$ \;
			\If{$score < bestscore$}{
				$best \leftarrow pos$ \;
				$bestscore \leftarrow score$ \;
			}
		}
		$L_i \leftarrow L_{i-1} \lhd_{pos} p_i$ \;
	} 
	\Return $L_n$\;
	\caption{\textsc{Docking}}
	\label{alg:dock}
\end{algorithm}

\begin{figure*}[t!]
\centering
  \includegraphics[width=\textwidth]{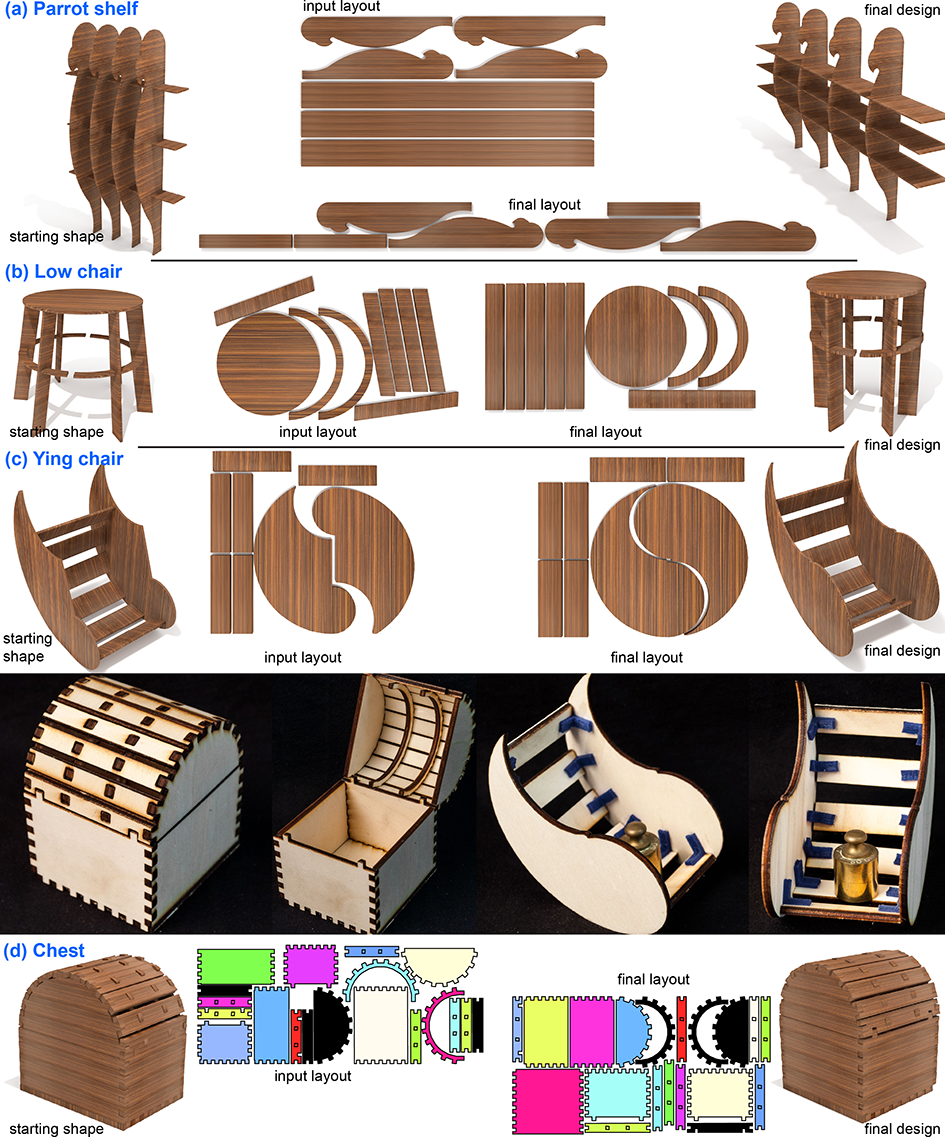}
\caption{Designs created using our system. Each design is shown with initial shape, starting layout, optimized layout, and final design. }
\label{fig:results_curved}
\vspace*{10mm}
\vnudge
\end{figure*}

\mypara{Docking criterion}
\label{sec:pack:crit}
The docking criterion considers wastage as the primary objective, where wastage
is defined by the ratio of occupied area divided by the bounding rectangle area
of the layout. We denote $W(L_{i})$ the wastage of a layout including up to part $i$.
It is obtained as $W(L_{i}) = \frac{\sum_{k=0}^i{A(p_k)}}{A(box(L_i))}$ where $A$ measures
area and $box(L)$ is the bounding rectangle of the layout. $W$ is therefore the ratio
between the area of the parts and the area of the bounding rectangle.

\begin{figure}[htb]
	\vnudge
	\centering
	\fbox{\includegraphics[width=0.45\columnwidth]{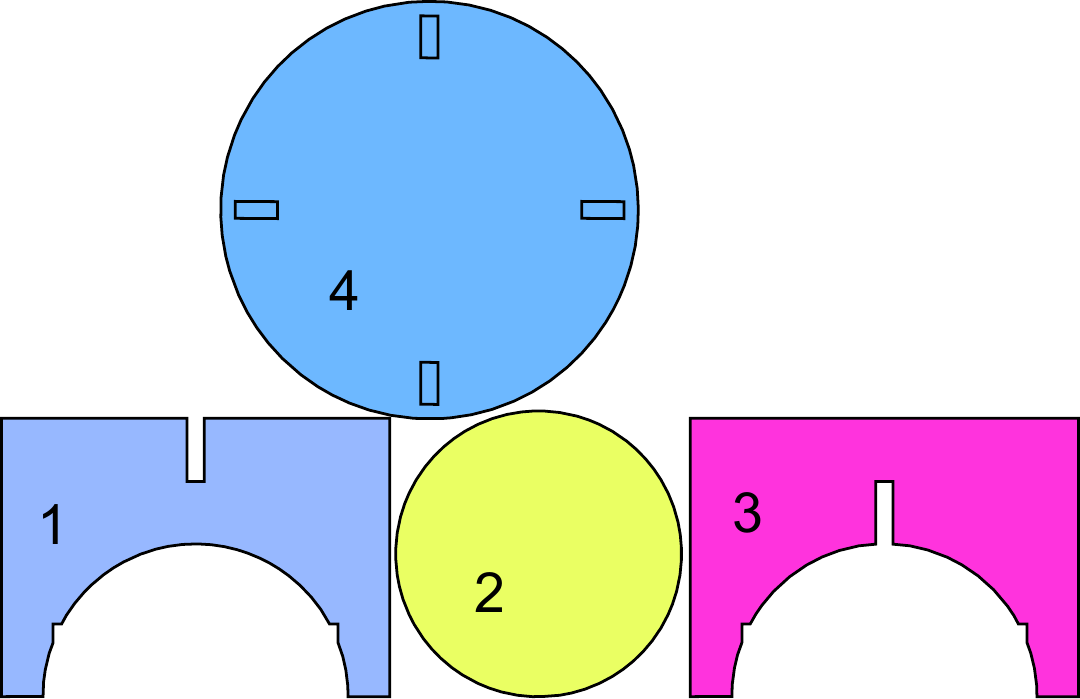}}
	\hfill	
	\fbox{\includegraphics[width=0.41\columnwidth]{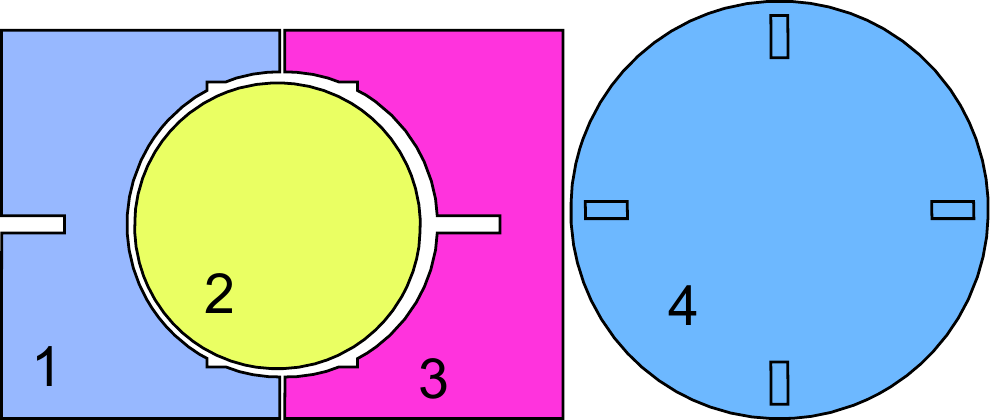}}
	\caption{Two layouts obtained with the same docking order. \textbf{Left:} Without taking enclosed area
		into account the first part is placed with the concavity against the bottom packing border. This prevents 
		the second part to nest within and cascades into a series of poor placements. \textbf{Right:} Taking
		into account enclosed areas results in a placement of the first part that allows nesting of the
		second part and produces a layout with lower wastage.
	}
	\label{fig:vsnaive}
\end{figure}

However, as the algorithm heuristically docks parts in sequence it cannot foresee
that some spaces will be definitely enclosed. In particular, for newly inserted
\textit{concave} parts there are often multiple orientations of the part resulting 
in the same wastage: if the concavity remains empty there is no preferred choice.
However, some choices are indeed better than others. If the concavity faces an
already placed object, then further docking \textit{within} the concavity 
will never be possible. This is illustrated in Figure~\ref{fig:vsnaive}, left.

We therefore propose a second criterion that discourages these bad choices.
The idea is to estimate the space that will be definitely enclosed when a part
is added to the current layout. This is done efficiently by considering
the enclosed space between the height-field of the current layout and the
height-field of the added part, along both horizontal and vertical directions.

Let $H^{r}(L)$ (respectively $H^{t}$) be the right (respectively top)
height-field of layout $L$ and $A(H^r(L))$ the area \textit{below} it. 
The enclosed area is then defined as:
\begin{equation*}
\begin{array}{l}
E(L_{i-1},p_i,pos) = \\[2mm]
\sum\limits_{s \in \{r,t\}} \max\left( 0, A(H^s(L_{i-1} \lhd_{pos} p_i)) - A(H^s(L_{i-1})) - A(p_i) \right) \\
\end{array}
\end{equation*}
with $A(p_i)$ the area of part $p_i$. Note the $\max$ that clamps negative values:
this is due to cases where the part nests in a concavity below the height-field
of the other direction.

The enclosed space is used as a tie-breaker when docking positions produce the same 
wastage values; therefore $D(L_{i-1},p_i,pos)$ returns the vector $( W(L_{i-1} \lhd_{pos} p_i), E(L_{i-1},p_i,pos) )$.
The effect of the enclosed area criterion is shown in Figure~\ref{fig:vsnaive}.

\begin{figure}[b!]
\centering
\includegraphics[width=.95\linewidth]{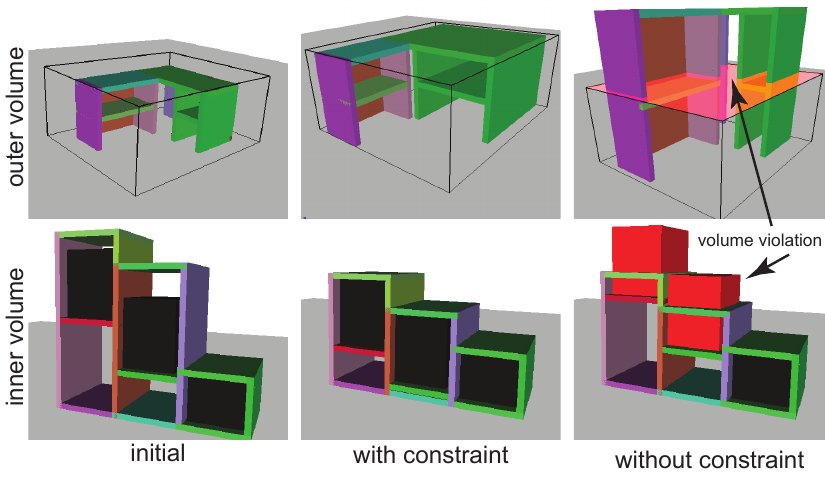}
\caption{We show effects of designing with~(middle column) or without~(right column) the respective constraints activated. }
\label{fig:designEfficiency}
\vnudge
\end{figure}

\section{Results}
\label{sec:results}

We used our system for various design explorations. As the complexity of the designs grows beyond 4-6 planks,  the utility of the system quickly becomes apparent. 
Note that the design constraints (see Figure~\ref{fig:designEfficiency}), by coupling different object parts, make the optimization challenging by preventing independent adaptation of part sizes. 
By off-loading  material usage considerations to the system, the user can focus on the design. Note that even when changes to the design are visually subtle, 
material utilization often increases significantly.  
%
%We allowed our colleagues (total of 6 computer science students) to try our system. After learning the design interface, they found the system very easy to use. The most time confusing part was proposing the initial design layout and specifying the design constraints. This part can probably be simplified by switching to a standard SketchUp like interface. We present a selection of complex examples created with our system
%in the results later in this section.

\mypara{Design examples}
We used our system to design and fabricate a range of examples comprising rectangular and/or curved parts. 
We fabricated fullscale and miniature models of designed furniture.
Models were made from MDF of 3~mm thickness and MDF of 30~mm thickness.
%, and pine wood of 20~mm thickness.
The designs are easy to manufacture in batches since after design layout optimization
they typically fit master boards completely: there is no need to attempt
to reuse leftover pieces of wood, and switching boards requires little clean up.
% Note that the weights for sagging were scaled down accordingly based on the plank thickness.

\begin{figure}[t!]
\centering
  \includegraphics[width=\columnwidth]{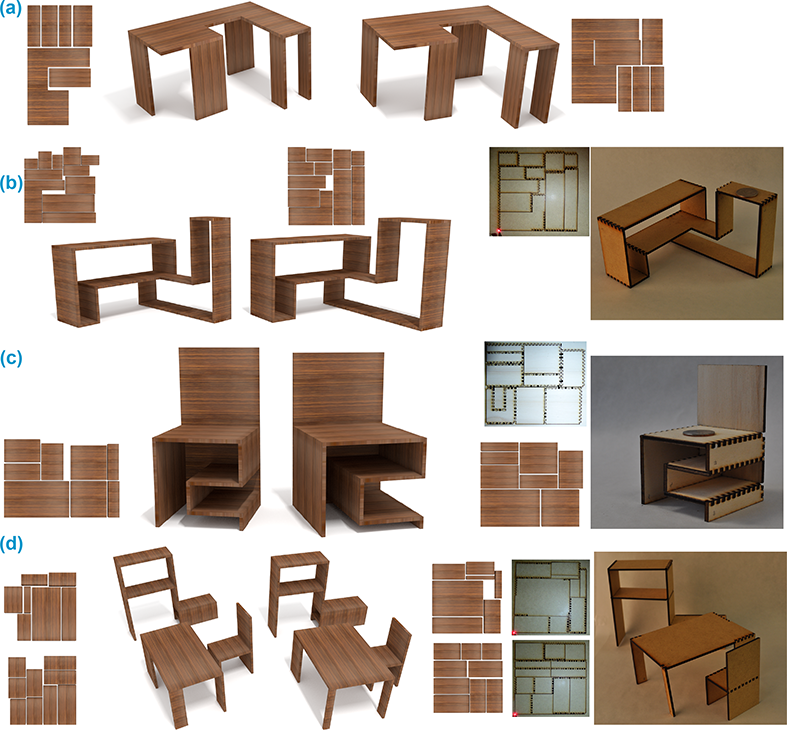}
\caption{Various material-driven design and fabrication examples. In each row, we show initial design (with material space layout inset), optimized design result (with material space layout inset), along with final
cutout  assembled model. Note that the design changes are often subtle, but still leads to significant improvement in material usage. }
\label{fig:results_plate_new}
\vnudge
\end{figure}

\begin{figure}[b!]
\centering
  \includegraphics[width=\columnwidth]{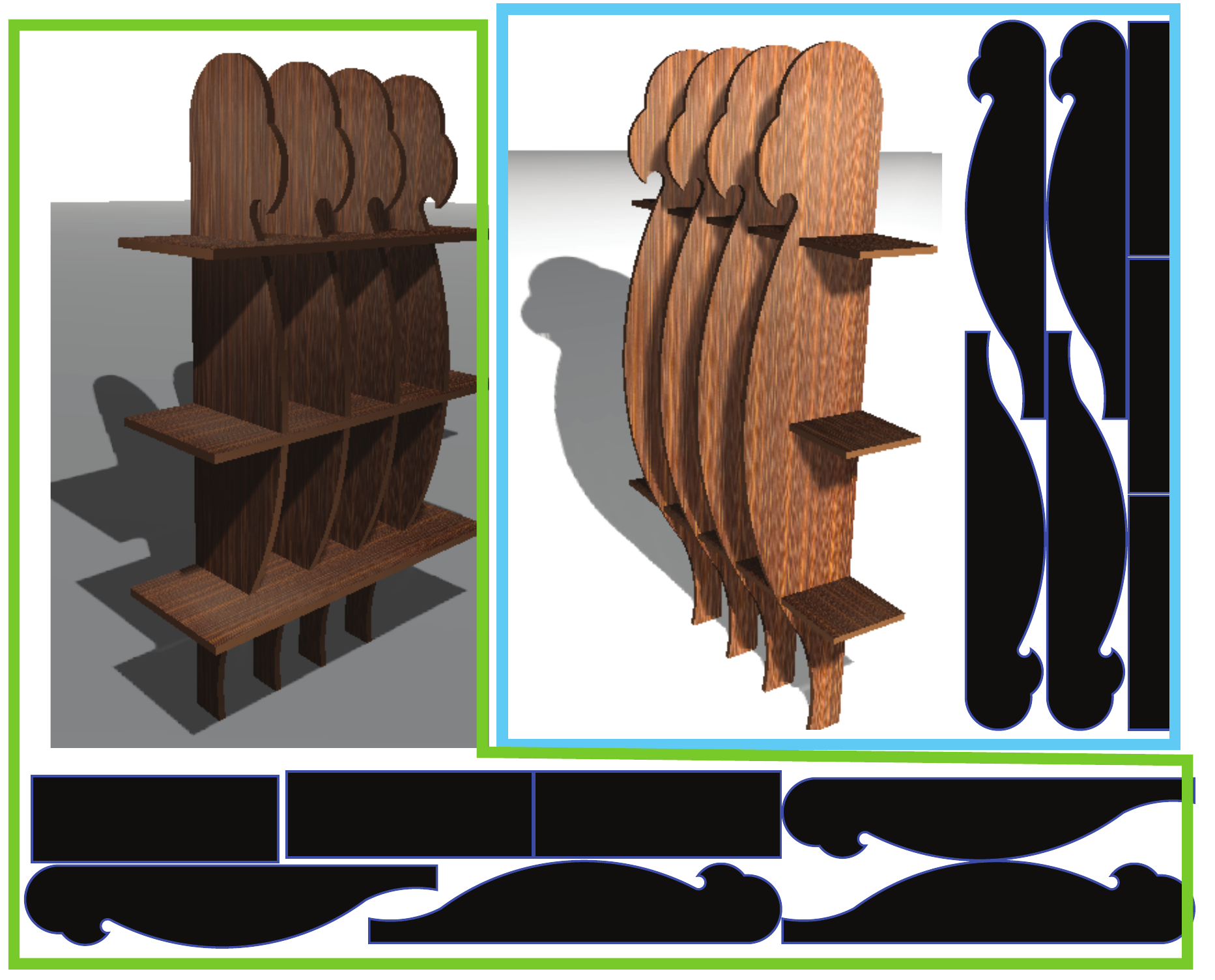}
\caption{Two different design suggestions (green has ratio 0.86, blue has ratio 0.85) for the parrot-shelf. Original design with another design suggestion is shown in Figure~\ref{fig:results_curved}. }
\label{fig:multiple_designs}
\vnudge
\end{figure}

We directly output the cutting plan for the laser cutter (or CNC machine) from the design layout,
adding connectors for planks sharing an edge, if needed. These are conveniently detected
since planks exactly overlap on edges in the 3D design. The connectors are either {\em finger
joints}, which are both strong after gluing and easy to assemble; {\em cross connectors}
for interleaved planks, or dowel-jointed for thicker materials (20~mm and 30~mm thickness).

\renewcommand{\arraystretch}{1.1}
\setlength{\tabcolsep}{.45em}
\begin{table}[t!]
\small
	\centering
	\caption{{\em Statistics for cut design showing the number of planks, number of constraints, material wastage ratio before and after the design suggestions/optimization.}}
	\begin{tabular}{l | c | c | c | c }
		& \#planks & \#constraints & ratio before & ratio after \\ \hline
		Figure~\ref{fig:teaser}                & 4  & 21 & 0.22 & 0.11  \\
		Figure~\ref{fig:results_curved}a & 7  & 33 & 0.34 & 0.08  \\
		Figure~\ref{fig:results_curved}b & 9  & N/A & 0.34 & 0.20  \\
		Figure~\ref{fig:results_curved}c & 8  & N/A  & 0.24 & 0.17  \\
		Figure~\ref{fig:results_curved}d & 16 & N/A  & 0.21 & 0.14  \\
		Figure~\ref{fig:results_plate_new}a & 6  & 22 & 0.15 & 0.04  \\
		Figure~\ref{fig:results_plate_new}b & 11 & 41 & 0.15 & 0.03  \\
		Figure~\ref{fig:results_plate_new}c & 8  & 13 & 0.26 & 0.03  \\
		Figure~\ref{fig:results_plate_new}d & 16 & 29 & 0.11 & 0.02  \\
	\end{tabular}
	\label{tab:stats}
\end{table}

%\begin{figure}[b!]
%\centering
%  \includegraphics[width=\linewidth]{figures/new/result_shelf.pdf}
%\caption{Optimized~(left) and fabricated~(right) shelf. Note that although the improvement in this case was $10\%+$, this is an easy example as large number of planks provide 
%extra design freedoms for the algorithm to explore.}
%\label{fig:results_shelf}
%\vnudge
%\end{figure}

Figures~\ref{fig:results_curved} and \ref{fig:results_plate_new} show various results. 
Table~\ref{tab:stats} gives an overview of the complexity of each model, and
the gains obtained by the layout optimizer. The system performs at interactive rates on a laptop taking from a few seconds to 3-4 minutes for the larger examples. Note that speed depends on how many exploration threads are pursued.

Figures~\ref{fig:teaser} and \ref{fig:results_curved} show results for objects with curved parts. Figure~\ref{fig:alog_steps} shows some intermediate shapes as the design evolves for the coffee-table (Figure~\ref{fig:teaser}) 
and the low-chair (Figures~\ref{fig:results_curved}-top) examples.  
Figure~\ref{fig:multiple_designs} shows alternate designs discovered by the algorithm for the Parrot shelf. While they have slightly lower usage they offer interesting variations that the user might prefer.

Figures~\ref{fig:teaser} was fabricated using a CNC machine. The optimized design achieved nearly $90\%$ material usage, although one can achieve null wastage by deciding to pick a rectangular top -- a decision that can be made after layout optimization as this opportunity is revealed. 
An allowable range was specified for the height and the bases were marked as symmetric as input design constraints.  
In the case of the parrot-shelf (Figure~\ref{fig:results_curved}a), the user indicated minimum and maximum range for the horizontal shelves along with desired range for the shelf heights. 

As described, parameteric designs are easily supported and optimized for in our framework. 
Figures~\ref{fig:results_curved}b-d show three such examples. In each case, additional constraints were provided to keep the objects within a given volume. The parts of the objects are all tightly coupled 
making these challenging examples to optimize for.

Figure~\ref{fig:results_plate_new}a shows a L-shaped work table. The user specified a target height for the design and a maximum work volume. Note that the legs of the table were also 
constrained to not change more than $25\%$ of original dimensions to prevent unwanted design changes. 
Figure~\ref{fig:results_plate_new}b shows a coupled shelf and table design where height of shelves and tabletop were similarly constrained. 
Figure~\ref{fig:results_plate_new}c shows a stylized chair, where both the chair seat height and chair width were constrained not to change beyond a margin. 
Figure~\ref{fig:results_plate_new}d shows multiple designs covering 2 master boards. 
The second master board is used as an overflow when docking can no longer fit a part in the first. The layouts are slid independently. 
%Note that the material utilization gets amplified for large scale manufacturing (e.g., Ikea), leading to significant
%reduction in production costs, both for material and for indirect costs such as
%cleaning, reuse and manipulation of the planks. 

%Figure~\ref{fig:results_shelf} shows design of a full sized shelf (span of 3m) was design and made out of 20mm pinewood slabs. As constraints, 
%the shelf heights were specified in an interval, and their was an upper bound on the maximum width of the shelf due to nearby pillars. Finally, to prevent sagging under load, each horizontal 
%plan was required to be of length 70cm or less (as recommended by http://www.woodbin.com/calcs/sagulator/). 

\begin{figure}[t!]
\centering
  \includegraphics[width=\linewidth]{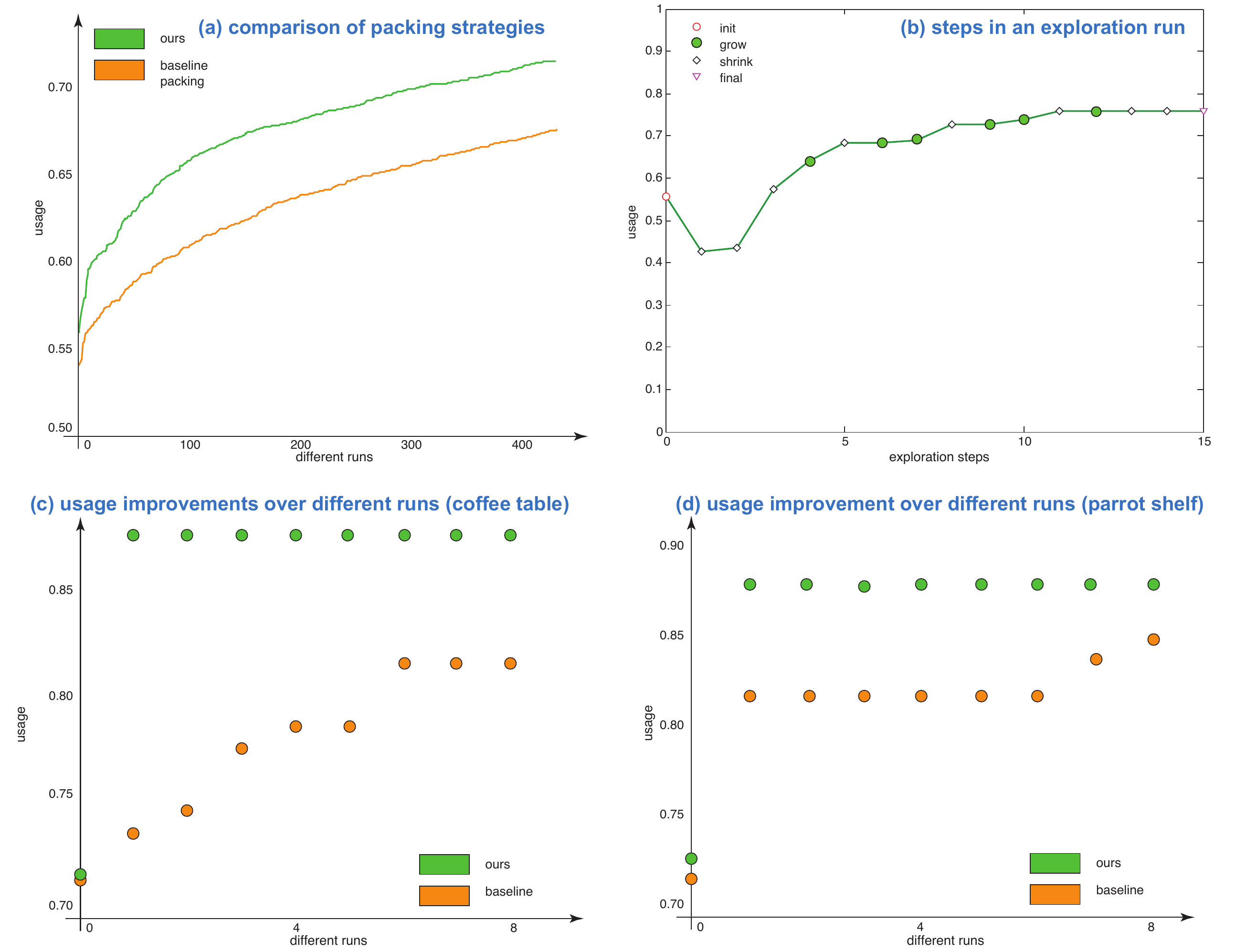}
\caption{Comparison of our algorithm against baseline alternatives. Higher is better. Please refer to the text for details.}
\label{fig:plot}
\vnudge
\end{figure}

\mypara{Comparison}
We now evaluate the relative importance of the key algorithm steps. 
Figure~\ref{fig:plot}a shows the importance of the docking criteria introduced in Section~\ref{sec:orderings}. We ran 500 random runs of our proposed packing algorithm with (`ours') and without (`baseline') the docking criteria on the coffee-table example. We sort the runs based on resultant usage (no 
shape optimization is performed here) and plot the two conditions. The docking criteria consistently resulted in $10$-$15\%$ better usage.
% Note that for only rectangular parts, this term will have no added value. \sylvain{actually it does -- it considers the 'shadow' cast by the part}
%

Figure~\ref{fig:plot}b shows usage improvement over one exploration run on the coffee-table sequence. The legend explains which step (grow, shrink, etc.) is being performed. 
While this is the result from a single thread, many similar threads are 
simultaneously explored. The few best results are then presented to the user as suggestions. 

Figure~\ref{fig:plot}c-d compare the importance of analyzing the material space layout to decide which plank to change and how. As baseline, we selected planks at random
and perform either a grow or shrink sequence with equal probability.
% and proposed parameter updates based on design space gradients.
Note that our method consistently outperforms the alternative approach.

\mypara{Design sessions}
We asked second year art students (6 subjects) from a design college to try our system. 
Figures~\ref{fig:results_plate_new}b-d show a selection of their designs. 
These particular students had performed a very similar task as part of their first year assignment -- `design furniture of your choice making best use of the provided piece of MDF board.' Hence, they were very aware of the
implicit link between design and material usage. Previously, they had used commercial 3D modeling tool (Rhinoceros, Solidworks, Sketchup Pro) for designing and mainly Illustrator for manually laying out the designs. They recalled the frustration of having to switch between the different 2D-3D design representations.
First, the students sketched design concepts before using our system.
Then, they used the exploration interface on their designs to reduce wastage.
Note that visually the initial sketch and final design can look similar, despite the increase in material utilization, which is desirable in terms of preserving the original design.

Overall, the feedback was positive. They appreciated being able to easily move between 2D$\leftrightarrow$3D, and not having to explicitly worry about material utilization.
%
%Most of them commented that material utilization was indeed one main concern for furniture design, especially for flatpack furniture. 
They appreciated the suggestions, instead of previous attempts using trial-and-error iterations between various softwares to reduce material wastage. %They sometimes used additional constraints so that the suggestion mode would not change important design parameters, e.g. the height of a table. 

\mypara{Limitations}
Currently, the algorithm can only make topological changes only for parameteric models. This will be an interesting future direction to pursue for constrained models. 
Our docking approach cannot nest parts into holes of other parts, a more advanced algorithm would be required.
A more material-induced restriction arises when the starting layout does not leave much space to optimize over. This effectively means that the degree of freedom for the design is low.
Adding more planks does reduce this problem (by providing additional freedom). However, beyond 25-30 planks, the exploration of the shape space becomes slow as there are too many paths to explore. One option is to limit exploration to only a subset of planks at a time, but then again, very desirable design configurations may be missed.
%
%In terms of compliance considerations, the grid-based treatment of element density can result in
%the optimizer preferring overlapping
%planks to accumulate twice the densities. This does not happen since overlapping
%two planks typically requires collapsing another plank below its minimal length, which we do not allow.
%Finally, we do not make any distinction between supporting a weight by an horizontal
%plank, or by a vertical plank going \textit{through} the weight. We avoid such
%issues by adding a constraint maintaining the weights vertically aligned
%on the horizontal plank supporting them, as specified by the user.

% ----------------------------------------------------------------
\section{Conclusions and Future Work}

We investigated how design constraints and material usage can be linked together towards form finding. Our system dynamically discovers and adapts to constraints arising due to current material usage, and computationally generates design variations to reduce material wastage.
By dynamically analyzing 2D material space layouts, we determine which and how to modify object parts, while 
using design constraints to determine how the proposed changes can be realized. This interplay results in a tight coupling between 3D design and 2D material usage and reveals information that usually remains largely invisible to the designers, and hence difficult to account for.
We used our system to generate a variety of shapes and demonstrated wastage reduction by $10\%$ to $15\%$.

Currently, we do not consider the stability of the produced furniture nor the durability
of the joints. This could be integrated as dynamic constraints following previous work on structural reinforcement~\cite{Stava:2012:SRI:2185520.2185544} and  
 shape
balancing~\cite{Prevost:2013:MSB:2461912.2461957}. % and furniture design~\cite{uim_guidedExploration_sigg12}.
Another important future direction is to generalize the framework to handle other types of laser cut materials, e.g., plastic plates that can be easily cut and more interestingly bend to have freeform shapes. Note that the packing problem will still be in 2D for such developable pieces. This can help produce interesting freeform shapes, while still making efficient use of materials.
%
%Finally, as a technical question, we would like to restrict suggested deformations to the null space of the stacked constraint matrix. The challenge here is to efficiently update the null space, without having to recompute the null space from scratch. This can result in  significant speedups.

\section*{Acknowledgements} The work was supported in part by ERC ShapeForge (StG-2012-307877) and ERC SmartGeometry (StG-2013-335373). 

% ----------------------------------------------------------------
% \input{fem.tex}

% ----------------------------------------------------------------
%\section*{Acknowledgements}

\bibliographystyle{acmsiggraph}
\bibliography{materialGuidedDesign}

% ----------------------------------------------------------------
%\input{additional_materials.tex}

\end{document}